\documentstyle[prd,aps]{revtex}
\begin{document}
\title{Analytical investigation for multiplicity difference correlators under
QGP phase transition}
\author{C.B. Yang$^{1,2}$ and X. Cai$^1$} 
\address{$^1$ Institute of 
Particle Physics, Hua-Zhong Normal University, 
Wuhan 430079, P.R. China\\
$^2$ Theory Division, RMKI, KFKI,
 Budapest 114., Pf. 49, H-1525 Hungary}
\date{ }
\maketitle

\begin{abstract}
It is suggested that the study of multiplicity difference correlators between two
well-separated bins in high-energy heavy-ion collisions can be used as a means to detect
evidence of a quark-hadron phase transition. Analytical expressions for the scaled
factorial moments of multiplicity difference distribution are obtained 
for small bin size with mean multiplicity ${\overline s}\le 0.3$ within
 Ginzburg-Landau description. It is shown that
the scaling behaviors between the moments are still valid, though the
behaviors of the moments with respect to the bin size are completely different from the
so-called intermittency patterns.
A universal exponent $\gamma$ is given 
to describe the dynamical fluctuations in the phase transition.
\end{abstract}


One of  the primary motivations of the study of high-energy heavy-ion collisions is to 
investigate the properties of quark-gluon system at extremely high temperature and high
density. Such system may be in the state of quark-gluon-plasma (QGP), and with the expanding 
and cooling the system will undergo a quark-hadron phase transition and turn out to be 
hadrons detected in experiments. One of the theoretical aims is to find a signal about
the phase transition.  As is well-known for a long time, 
fluctuations are large for statistical systems near their critical points.
Thus the study of fluctuations in the process might reveal some features
for the phase transition.$^{[1,2]}$ Monte Carlo simulations $^{[3]}$ on intermittency$^{[4]}$
without phase transition for $pp$ collisions $^{[5]}$ show quantitatively different results
on multiplicity fluctuations from theoretical predictions with the onset of phase
transition.$^{[2]}$ These different results stimulated a lot of theoretical works
on multiplicity 
fluctuations with phase transition of second-order $^{[2,6]}$ and first-order $^{[7,8]}$ 
within Ginzburg-Landau model which is suitable for the study of phase transition for 
macroscopic systems. Most of these works give remarkable scaling behaviors between 
 $F_q$ and $F_2$, and there seems to exist a universal exponent $\nu$ $^{[2,6,8]}$. It is suggested that the exponent $\nu$ can be used as a useful 
diagnostic tool to detect the formation of QGP. In [7] $\ln F_q$ are expanded
as power series of $\delta^{1/3}$, and it is shown that the coefficient of 
$\delta^{1/3}$ term can be used as a criterion for the onset and the order of the phase
transition. All those works show the violation of intermittency patterns in the
phase transition.

It is known for a long time that the investigation of multiplicity fluctuations
is very different in heavy-ion collisions, though the power-law dependence
of $\ln F_q$ on $\delta$, $F_q\propto \delta^{-\varphi_q}$, has been found
ubiquitous in hadronic an leptonic processes.$^{[9]}$ The main differences
between heavy-ion physics and hadronic \& leptonic ones
on multiplicity fluctuations were noticed earlier in Ref. [10]. In Ref. [11] 
an alternative way was proposed to study the fluctuations by means of
factorial moments of the multiplicity difference (FMMD) between two well-separated 
bins. This alternative is a hybrid of the usual factorial correlators $^{[4]}$ and
wavelets $^{[12]}$ because $W_{jk}$ in Haar wavelet analysis is just the difference of
multiplicities in two nearest bins. Present discussions, of course, will not
be limited in the nearest bins. Let the two bins, each of size $\delta^2$
and separated by $\Delta$, have multiplicities $n_1$ and $n_2$, and define their
multiplicity difference $m=|n_1-n_2|$. Scaled FMMD are defined as
\begin{equation}
{\cal F}_q=f_q/f_1^q, \mbox{\hspace{0.8cm}}
f_q=\sum_m m(m-1)\cdots (m-q+1) Q_m\ ,
\end{equation}

\noindent with $Q_m$ the distribution of multiplicity difference which may be
dependent on $\Delta, \delta$ and details of the process. Moments defined 
above are similar to but not the same as the Bialas-Peschanski correlators
$^{[4]}$ $F_{q_1q_2}$, for ${\cal F}_q$ may depend on both $\Delta$
and $\delta$ while $F_{q_1q_2}$ depends only on $\Delta$. 

In Ref. [11], ${\cal F}_q$ are numerically studied within
Ginzburg-Landau model. The scaling behaviors between $F_q$ and $F_2$, 
${\cal F}_q\propto {\cal F}_2^{\beta_q}$, are shown with
$\beta_q=(q-1)^\gamma$ and a universal exponent $\gamma$=1.099.

In this paper, ${\cal F}_q$ are studied analytically for very small bin size
$\delta$. Then the dynamical fluctuation components ${\cal F}^{\rm (dyn)}_q$
of FMMD are defined. It is shown that
both $\ln {\cal F}_q$ and $\ln {\cal F}^{\rm (dyn)}_q$ increase linearly
with the bin size $\delta$ when $\delta$ is very small, completely different
from the usual intermittency behaviors of $\ln F_q$ which increase with the
decrease of bin size. But the scaling laws
between ${\cal F}_q$ and ${\cal F}_2$, and between ${\cal F}_q^{\rm (dyn)}$
and ${\cal F}_2^{\rm  (dyn)}$
are still valid, although the corresponding $\beta_q$ and $\beta_q^{\rm (dyn)}$
are different. A universal exponent $\gamma$ for $\beta_q^{\rm (dyn)}$
is given which has no dependence on any parameter in the model
and is different from that in [11]. 
  
As an starting point, let us first discuss the trivial and simplest case.
Suppose that the two bins considered are well-separated so that there is no
correlations between them. Let the mean multiplicities in each bin are $s_1,
s_2$, respectively. If there is no dynamical reason, the multiplicity
distribution for each bin is a Poisson one
\begin{equation}
P_{n_i}(s_i)={s_i^{n_i}\over n_i!} \exp(-s_i)\ \ (i=1,2)\ .
\end{equation}

\noindent From this distribution, one can deduce the multiplicity difference
distribution as
\begin{equation}
P_m(s_1, s_2)=\cosh({m\over 2}\ln {s_1\over s_2}) I_m(2\sqrt{s_1s_2})e^{-s_1-s_2}\ (2-
\delta_{m0})\ ,
\end{equation}

\noindent where $I_m(z)$ is the modified Bessel function of order $m$,
\begin{eqnarray*}
I_m(z)=\sum_{k=0}^{\infty} {(z/2)^{2k+m}\over k!(k+m)!}\ .
\end{eqnarray*}

\noindent FMMD for pure statistical fluctuations are
\begin{equation}
f_q^{\rm (stat)}=\sum_{m\ge q} m(m-1)\cdots (m-q+1) P_m(s_1, s_2)\ .
\end{equation}

For simplicity, let us discuss the case with $s_1=s_2$. This condition can
always be satisfied if one chooses the two bins properly.
Since only $m\ge q$ contribute to
$f_q$, the summation over $m$ in last equation  can be extended to $m=0$.
This summation converges very slowly because $I_m(2s)$ decreases with $m$
approximately as $s^m/m!$ for large $m$ and small $s$, but the product
$m(m-1)\cdots (m-q+1)$ increases with $m$ quickly. 
So contributions from all $m\ge q$
must be taken into account. This will cause some difficulties in numerical
calculations if one starts directly from the definition of the moments.
In this paper, we will alternatively sum over $m$ analytically, and then do 
numerical calculations from the final expression. In this approach, one can
control the precision more easily in calculation. To complete the summation, one
can introduce a generating function
\begin{equation}
G(x,s)=2e^{-2s}\sum_{m=0}^\infty x^m I_m(2s)\ ,\mbox{\hskip 0.5cm}
G_q(x, s)={d^q G(x,s)\over dx^q} \ .
\end{equation}

\noindent With this function, $f_q^{\rm (stat)}$ can be rewritten as
\begin{equation}
f_q^{\rm (stat)}=G_q(1,s)\equiv G_q(s)\ .
\end{equation}

\noindent Direct algebra shows that
\begin{equation}
G(x,s)=2e^{(x-2)s}\left[a_0^0+\sum_{i=1}^\infty a_i^0 {d^i\over dx^i}{1-\exp(-xs)
\over x}\right]
\end{equation}
 
\noindent with $a_i^0=(-1)^i s^{2i}/(i!)^2$ for $i=0, 1, \cdots$, and that
\begin{equation}
f_q^{\rm (stat)}=2e^{-s}\left[a_0^q+\sum_{i=1}^\infty a_i^q\sum_{j=i}^\infty
{(-s)^{j+1}\over j(j-i)!}\right]\ ,
\end{equation}

\noindent where $a_i^q$ can be calculated by recurrence relation from
$a_i^0$, $a_0^q=sa_0^{q-1}$, $a_1^q=sa_1^{q-1}$, $a_i^q=sa_i^{q-1}+
a_{i-1}^{q-1}\ , (i\ge 2)$. Then one can get two specially important
coefficients $a_0^q=s^q$, $a_1^q=-s^{q+2}$. The most important advantage
of such calculations is that these formalisms facilitate analytical
calculations for quantities  in the range of very small bin size in which
we are now interested. We will discuss it later in this paper.

Now, we begin to discuss the FMMD 
in second-order quark-hadron phase transition. We use the Ginzburg-Landau description
to specify the probability that $s$ hadrons are created in the two dimensional, such as
$\delta\eta\delta\varphi$, area $\delta^2$. In this description, the distribution of 
multiplicity is no longer a Poisson one and that for multiplicity
difference is given by$^{[11]}$
\begin{equation}
Q_m(\delta, \tau)=Z^{-1}\int {\cal D}\phi P_m(\delta^2\tau\mid\phi\mid^2)e^{-F[\phi]}\ ,
\end{equation}

\noindent where $\tau$ is an indication of lifetime of the whole parton system,
${\cal D}\phi=\pi d\mid\phi\mid^2, Z=\int {\cal D}\phi e^{-F[\phi]}$ and
the free energy $F[\phi]=\int_{\delta^2} dz\left[a\mid\phi\mid^2+b\mid\phi\mid^4
+c\mid \partial \phi/\partial z\mid^2\right].$

As has been pointed out in [2, 6] that for small bin the gradient term in $F[\phi]$
does not have any significant effect on the multiplicity fluctuation, so one can
set $c=0$. This setting means that $\phi$ can be regarded as a constant over the area
$\delta^2$. Of course, this is approximately true only when $\delta^2$ is very small.

Substituting $Q_m(\delta, \tau)$ into Eq. (1), one gets
\begin{equation}
f_q=\left.\int_0^\infty du\, G_q(\tau xu)\,e^{xu-u^2}\right/\int_0^{\infty}
du\, e^{xu-u^2}
\end{equation}

\noindent with $x=\mid a\mid\delta/b$ related with the bin width $\delta$. 
Define$^{[6]}$
\begin{equation}
J_q(\alpha)=\int_0^{\infty}\ du\  u^q\ e^{\alpha u-u^2}
\end{equation}

\noindent which satisfies recurrence relation $J_q(\alpha)={\alpha\over 2} J_{q-1}(\alpha)
+{q-1\over 2} J_{q-2}(\alpha)$ and can be directly integrated for $q=0$ and 1,
$J_0(\alpha)={\sqrt{\pi}\over 2}e^{\alpha^2/4}(1+{\rm erf}({\alpha\over 2}))$,
$J_1(\alpha)={1\over 2}+{\alpha\over 2}\,J_0(\alpha)$. With $J_q(\alpha)$,
$f_q$ can be expressed as
\begin{equation}
f_q=J_0^{-1}(x)\sum_{i=q}^\infty b_i^q\,(\tau x)^i J_i(-(\tau-1)x)
\end{equation}

\noindent with $b_i^q$ constants, $b_q^q=1$,especially. Notice that the second nonzero
$b_i^q$ for fixed $q$ is for $i=q+4$. One can check this from the expression
for $G(x,s)$ and the recurrence relations for $a_i^q$.

The scaled FMMD ${\cal F}_q$ defined do contain contributions from statistical
fluctuations, contrary to the usual scaled factorial ones. As a way to seek for
the dynamical fluctuations, one can define the dynamical scaled FMMD as
\begin{equation}
{\cal F}_q^{(\rm dyn)}={{\cal F}_q\over {\cal F}_q^{\rm (stat)}}\ .
\end{equation}

\noindent To make the definition sense, one should ensure that the mean multiplicity 
is the same
for all the calculation of the moments concerned. In Ginzburg-Landau model, the mean 
multiplicity is ${\overline s}=\tau x\,J_1(x)/J_0(x)$. Then deviations of ${\cal
F}_q^{(\rm dyn)}$ from one should indicate the existence of dynamical fluctuations.
 The three classes of moments defined in this paper can all be calculated directly.

Up to now, all of moments are expressed as infinite sums and are exact within the
model. The infinite summation will hinder us from an explicit
formalism for interesting quantities. Now we focus on the range of very small bin size.
As has been shown, the smallness of the bin size $\delta$ is for the need of 
self-consistence of Ginzburg-Landau model adopted in this paper, 
otherwise the gradient term
plays a role and cannot be set to zero. Experimentally, the bin size $\delta$ can be
chosen very small indeed. For example, 
experimental data$^{[13]}$ show that  the number of total produced charged particles is about 
70 within a rapidity range about 7 in 200 $A$ GeV S+Em collisions.
The rapidity resolution in EMU01 experiments can be 
high up to 0.01. In two dimensional analysis as in this paper, the area $\delta^2$ 
considered can be so small that in that area the mean multiplicity satisfies $s\ll 1$.
The mean multiplicity in single bin can still be much less than 1 even for
Pb-Pb collisions in which the number of produced particles can reach 1500
or more.
Because of such experimental facts, we can discuss
only the cases with $s\ll 1$ in the following, and our results can be checked directly
in experiments. One can see that this condition 
will enable us to reach simple expressions for all the moments.

For the pure statistical fluctuation case, terms excepts the leading term in Eq. (8)
can be neglected, and one can easily get
\begin{equation}
\ln F_q^{\rm (stat)}=(q-1)({\overline s}-\ln 2)\ .
\end{equation} 

\noindent One can check that the relative contribution from all non-leading
terms is about 1\% for ${\overline s}=0.3$.
 
For the moments with the onset of phase transition, it is a little complicated
because of the integration in Eq. (10) over the whole range of $s$. But, one can 
see that the leading term in $G_q(s)$ plays a dominated role. One needs to notice 
that integrating $u^n$ term is associated with product of two factors $(\tau x)^n$
and $\exp(-\tau^\prime xu-u^2),\ \tau^\prime=\tau-1$. For small $x\tau$ 
the first factor strongly suppresses the contribution. For larger $x\tau$ the term 
 $\exp(-\tau^\prime x u)$ over-depresses the contribution from the former. 
In fact, numerical results show that $(x\tau)^4 J_{q+4}
(-\tau^\prime x)/J_q(-\tau^\prime x)$ is always of the order $10^{-4}$ for
$x\tau\le 0.5$, corresponding to ${\overline s}\simeq 0.3$ for $\tau=10.0$.
So that the results will not be affected practically if only the leading
term are kept for the calculation of the moments in small $x$ region. 
Then to a good approximation,
\begin{eqnarray}
\ln {\cal F}_q& = & (q-1)\ln {J_0(x)\over 2}+\ln J_q(-\tau^\prime x)-q\ln J_1(-\tau^\prime x)\ ,\\
\ln {\cal F}_q^{\rm (dyn)}& = & (q-1)\ln{J_0(x)\over \exp({\overline s})}+\ln J_q(-\tau^\prime x)
-q\ln J_1(-\tau^\prime x)\ .
\end{eqnarray}

The behaviors of $\ln {\cal F}_q$ and $\ln {\cal F}_q^{\rm (dyn)}$ as functions
of $x$ from 0.005 to 0.05 are shown in Fig. 1 for 
$\tau$=2.0 and 10.0. The $x$ range
is chosen from the requirement ${\overline s}\le 0.3$ for $\tau$=10.0. One can see
that both $\ln {\cal F}_q$ and $\ln {\cal F}_q^{\rm (dyn)}$ have linear dependence
on bin size $x$. This dependence is completely different from the usual intermittency
behaviors. This result can also
be seen directly from last expressions for the moments if one substitutes 
$J_q(\alpha)$ with ${1\over 2}\left(\Gamma({q+1\over 2})+\alpha\Gamma({q+2\over 2})
\right)$ for very small $\alpha$. In small $x$ approximation,
\begin{eqnarray}
&&\ln {\cal F}_q = {\rm const}+\left[ (q-1){\Gamma(1)\over\Gamma({1\over 2})}-\tau^\prime
\left({\Gamma({q+2\over 2})\over \Gamma({q+1\over 2})}-q{\Gamma({3\over 2})\over
\Gamma(1)}\right)\right]x+O(x^2)\ ,\\
&&\ln {\cal F}_q^{\rm (dyn)} = {\rm const}+\tau^\prime \left[q{\Gamma({3\over 2})\over
\Gamma(1)}-{\Gamma({q+2\over 2})\over \Gamma({q+1\over 2})}-(q-1){\Gamma(1)\over
\Gamma({1\over 2})}\right]x+O(x^2)\ .
\end{eqnarray}

\noindent Numerical results show trivial scaling behaviors for $\ln {\cal F}_q$ vs
$\ln {\cal F}_2$ and $\ln {\cal F}_q^{\rm (dyn)}$ vs $\ln {\cal F}_2^{\rm (dyn)}$
in Fig. 2. Though $\ln {\cal F}_q$ have different ranges of values for different
$\tau$, the scaling behaviors seem independent of the lifetime of the system.
One can see weak dependence on $\tau$ for $\beta_q$
from last equations. $\beta_q^{\rm (dyn)}$ do not depend on any parameter
in the model because the $\tau$ dependencies in the local slopes are cancelled 
miraculously with each other in small $x$ limit. More interestingly,
$\beta_q^{\rm (dyn)}$ can be well fitted by
\begin{equation}
\beta_q^{\rm (dyn)}=(q-1)^\gamma
\end{equation}

\noindent with $\gamma$=1.3424, as shown in Fig. 3. But $\beta_q$ do not obey the same
scaling law, as shown in Fig. 3 for the case with $\tau=10.0$. The universal exponent
$\gamma$ is different
from that in [11]. But the difference does not mean any contradiction between
present paper and [11], because they correspond to different quantities. The 
difference also comes from the different $x$ regions discussed since $\beta_q$ 
depend on the fitting range. In this paper, the exponent $\gamma$ is completely
determined by the general features but does not depend on any
parameter of Ginzburg-Landau model
used to describe the phase transition. The exponent $\gamma$ given here is very close
to the exponent $\nu$ given in former studies on multiplicity fluctuations for
second-order phase transition. The slight difference between them comes from the different
regions concerned. As shown in Fig. 3 of the first paper in [6], $\beta_q$ take minima at
$\alpha\equiv(x/2)^{0.5}\simeq 1$ and increase with the decrease of $\alpha$. Our result
corresponds to $\alpha=0$. Since both $\nu$ and $\gamma$ describe dynamical
fluctuations in phase transition, they should be equal, as physically demanded.
For experiments with ${\overline s}$ in single bin a little larger than 0.3
the experimentally obtained $\gamma$ should be close to but less than
1.3424. Thus if experiments observe a scaling
exponent $\gamma$ about 1.34 in a high resolution analysis, the onset of a
second-order quark-hadron phase transition can be pronounced.

It should be pointed out that even without dropping off non-leading terms,
the exponent $\gamma$ in this paper will not be changed, because all those terms
are related to higher orders of $x$ and have no contribution to $\gamma$ which
is connected with properties of the moments in the limit $x\to 0$. In this sense,
the exponent $\gamma$ given here is exact and truly universal.

In summary, scaled FMMD are studied analytically within Ginzburg-Landau model
in a kinetical region with mean multiplicity in single bin less than 0.3 for
second-order quark-hadron phase transition. The dynamical fluctuations in FMMD
are extracted, which give the same physical contents as the usual scaled factorial
moments. Scaling behaviors between scaled FMMD are shown, and a truly universal
exponent is given.

This work was supported in part by the NNSF, the SECF and Hubei NSF in China.
One of the authors (C.B. Yang) is grateful for fruitful discussions with
Prof. R.C. Hwa and would like to thank Prof. T.S. Bir\'o and Prof. P. L\'evai
for kind hospitality during his stay in Hungary.

\centerline{{\Large Figure Captions}}
\begin{description}
\item
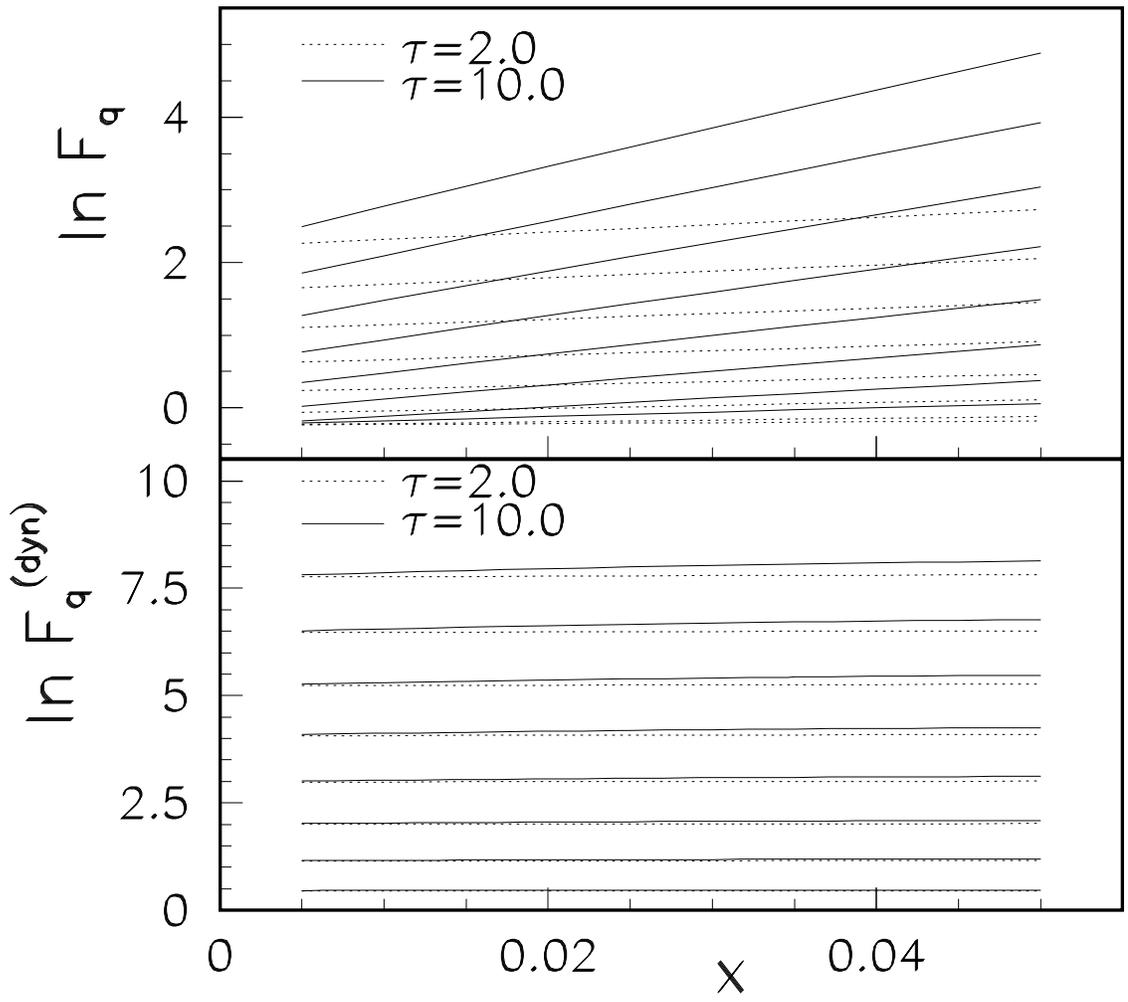
{\bf Fig. 1} Dependences of $\ln {\cal F}_q$ and $\ln {\cal F}_q^{\rm (dyn)}$
on the bin width $x$ for $\tau=10.0$ and $\tau=2.0$.
\item
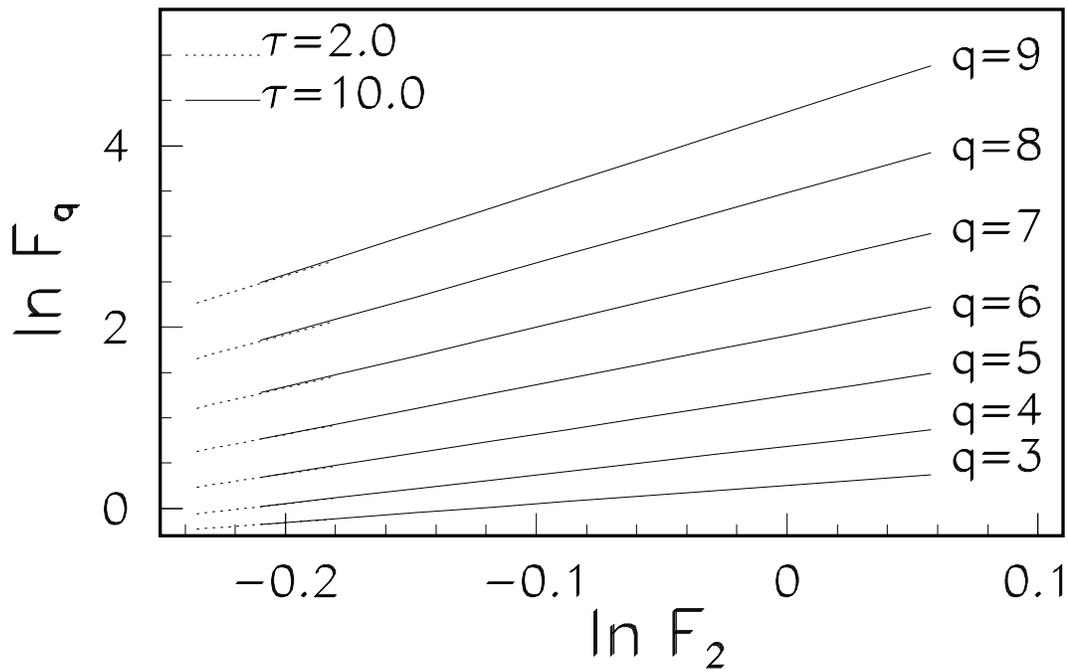
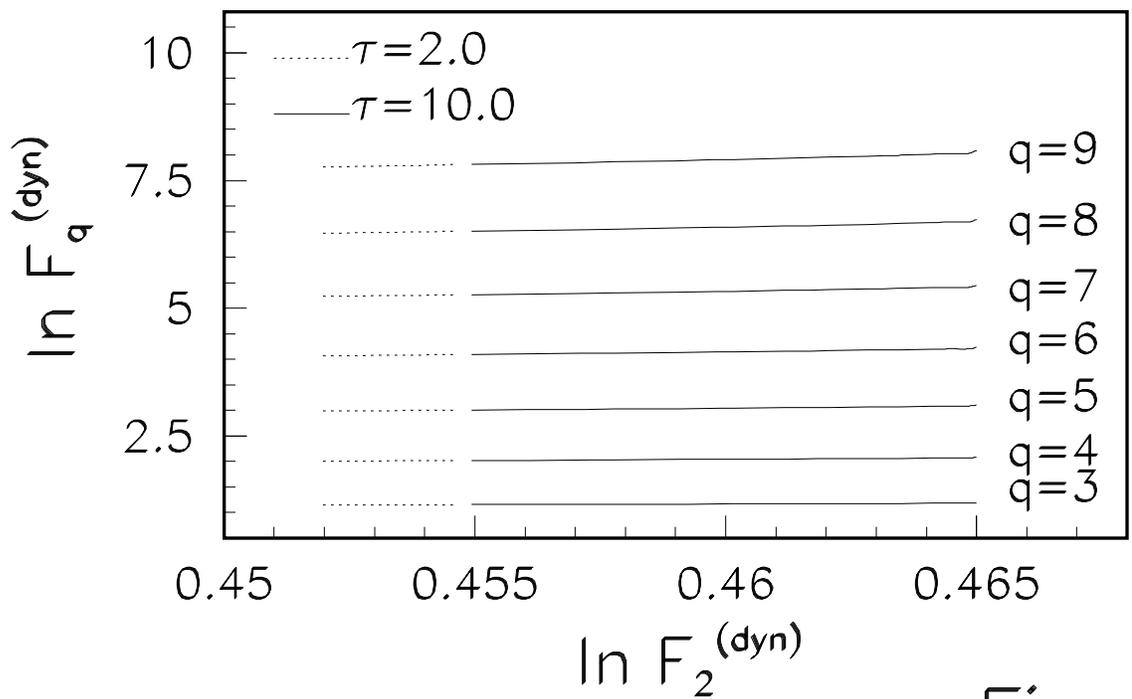
{\bf Fig. 2} Scaling behaviors of $\ln {\cal F}_q, \ln {\cal F}_q^{\rm (dyn)}$ vs
$\ln {\cal F}_2, \ln {\cal F}_2^{\rm (dyn)}$ for the same choices of lifetime as in Fig.1.
\item
{\bf Fig. 3} Scaling exponent $\ln \beta_q$ vs $\ln (q-1)$.
\end{description}
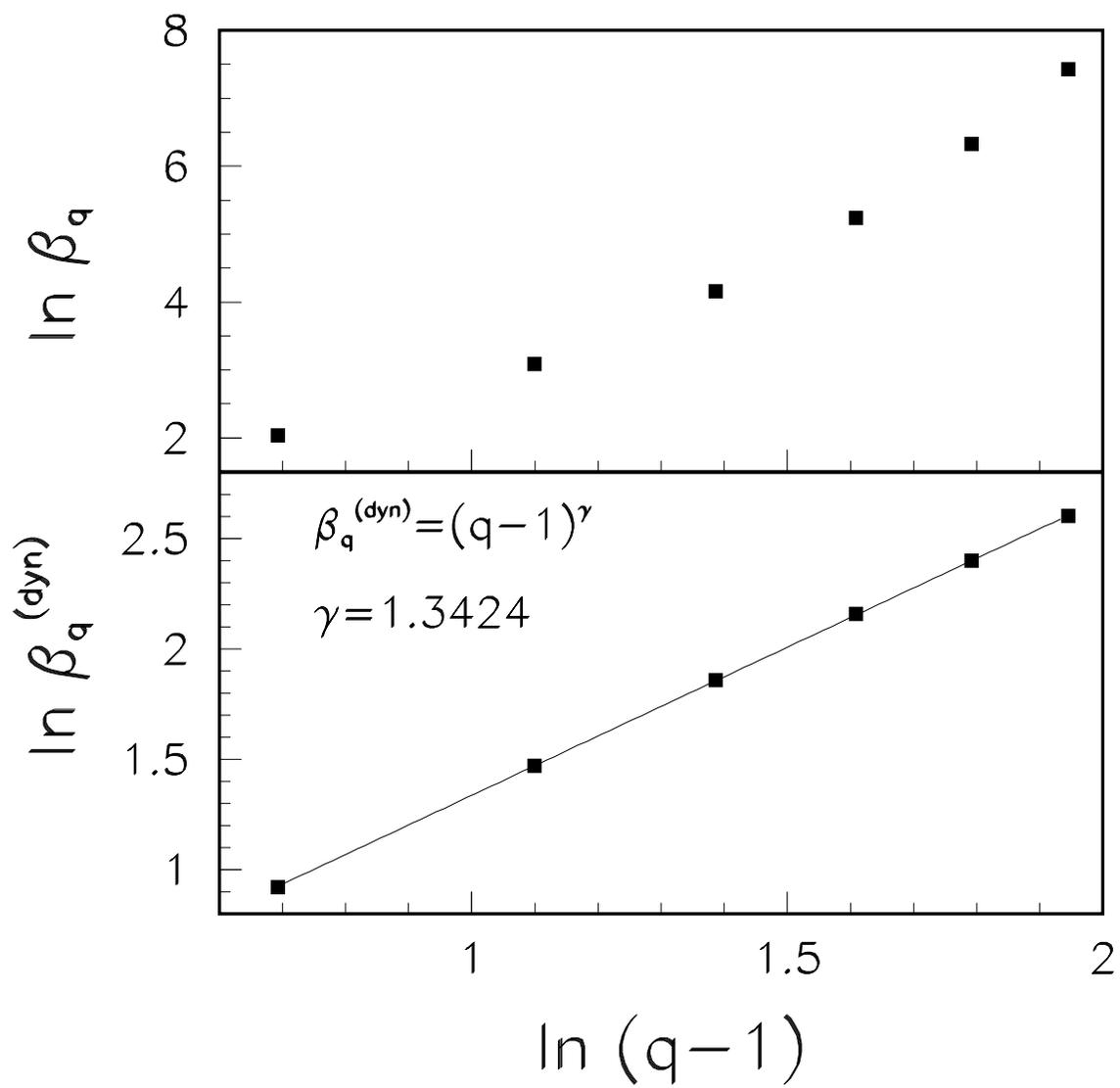
\end{document}